\documentclass{llncs}
\usepackage{graphicx}
\usepackage{hyperref} 
\usepackage{subfig}

\begin{document}
\pagestyle{headings}  
\title{An applied spatial agent-based model of administrative boundaries using SEAL}
\author{Bernardo Alves Furtado \inst{1,2} \and Isaque Daniel Eberhardt Rocha\inst{1}}

\institute{Institute of Applied Economic Research, Brazil\\
\email{bernardo.furtado@ipea.gov.br}\\ 
\and
National Council of Research, CNPq}

\maketitle             
\begin{abstract}
This paper extends and adapts an existing abstract model \cite{furtado_jasss} into an empirical metropolitan region in Brazil. The model - named SEAL: a Spatial Economic Agent-based Lab - comprehends a framework to enable public policy \textit{ex-ante} analysis. The aim of the model is to use official data and municipalities spatial boundaries to allow for policy experimentation. The current version considers three markets: housing,  labor and goods. Families' members age, consume, join the labor market and trade houses. A single consumption tax is collected by municipalities that invest back into quality of life improvements. We test whether a single metropolitan government - which is an aggregation of municipalities - would be in the best interest of its citizens. Results for 20 simulation runs indicate that it may be the case. Future developments include improving performance to enable running of higher percentage of the population and a number of runs that make the model more robust as well as further policy analysis.
 
\keywords{Agent-based model, Public Policy, Spatial modeling, Metropolitan regions, Public Finances}
\end{abstract}

\section{Introduction}

Anticipating public policy effects is no simple task \cite{colander_kuper,geyer_cairney,furtado_modeling_book}. Once planned and put into effect, policies are subject to interested actors reaction, learning, chance, and emergent behavior \cite{miller_page}. That is why governments actions are monitored and \textit{ex-post} evaluation is implemented.

Understanding of policies' effects \textit{ex-ante} is relevant in the Brazilian context for at least two main reasons. Firstly, municipalities in Brazil are autonomous, independent entities, comprising a three-level federation since the 1988's Constitution. That implies metropolitan neighbors can act politically independently in terms of transportation or other public services.

Secondly, metropolises in Brazil tend to be centralized with concentration of economic activity, higher level of business and amenities in the capital. Thus, agglomeration economies and fiscal regulation favor capital cities in detriment of nearby municipalities \cite{furtado_finances}. Peripheral areas, in turn, observe higher levels of violence, longer commuting times to the center and lower quality of public services offer. In practice, that means most workers commute to the capital, contribute to increased economic output, but go home at night to an outer municipality that does not collect enough taxes to supply basic services. 

As a result, urban economic space in Brazil is highly heterogeneous with concentration of capital and quality in the center and lack of access to services and jobs in the borders. Such spatial configuration, when divided by political-administrative boundaries, tend to further segregate access to citizens and, thus, lower economic output. In fact, political fragmentation in metropolises has been found to reduce productivity \cite{ahrend}. 

Given such a context for the Brazilian case, an exercise of \textit{ex-ante} policy analysis that could verify whether a political entity spatial configuration serves the public best may be helpful. Nevertheless, there is no evidence of such exercise. Probably because solving what-if problems in an urban environment looking towards the future is methodologically burdensome. 

However, recent developments in agent-based modeling, especially for economic \cite{dawid_2014,dosi_2012}  and urban analysis \cite{filatova_2009}, have been shown to contribute to the debate and enlighten future prospects of policy analysis. Moreover, ABMs are helpful in modeling complex systems such as markets \cite{dawid2,boero} in which interactions among agents are not linearly related to observed macro structures. 

The model presented here is intended as a simple one \footnote{There has been a debate in ABM literature about whether a model should be large and complex such as \cite{derhoog,guocheng} or simple, as argued by \cite{lengnick}.} in which the main behaviors of the economy in space are captured. Hence, we depart from an abstract previous model \cite{furtado_jasss}, turn the model into an empirical one and test a specific policy, i.e., whether bounded municipalities, with a single metropolitan government are more beneficial to the average citizen. Such benefit is tested using a Quality of Life Index (QLI) weighted by the population. This test is provided as an illustration of ex-ante policy analysis. Other possibilities of the model as a framework are also discussed in \ref{final}.

The main contribution of the paper is an extension and testing of a simple, fully functional, spatial modeling that is based on previous work \cite{lengnick,gaffeo,furtado_jasss}. Yet, it goes further than land-use change analysis \cite{filatova_2009,parker} to include spatially governmental entities and fiscal testing in an applied exercise. Thus, moving the arguments of \cite{geyer_cairney} and \cite{colander_kuper} into a practical application.

Besides this introduction, the next section (\ref{model}) presents and details the model and the agents' behavior. Then preliminary results are discussed (\ref{results}) along with the limitations of this version. The closing section (\ref{final}) develops the scope of the model and details further developments.

\section{Model and processes}\label{model}

The objective of SEAL is to model economic markets in space and time so that policies and budgetary decisions can be evaluated given actual municipal boundaries. Thus, citizens reside in domiciles that are precisely located in urban and rural space within different municipalities. Businesses also have geographical coordinates. These locations allow the model to implement decisions based on distance, such as shopping at the nearest shops or working closest to home. Also, production and tax collections are localized. That means the spatial design of taxes can be investigated. This is specially relevant for the case of Brazilian metropolises in which there is clear disagreement among urban functional areas and offer of public services \cite{furtado_finances} and commuting and transport are relevant given the poor quality of infrastructure.

The model in \cite{furtado_jasss} is based on \cite{lengnick} and \cite{gaffeo}. However, the original abstract model is built upon seven, squared and fictitious regions with 1,000 agents in total. The one in this paper is based on actual municipal boundaries of the metropolitan region of Bras\'{i}lia, in Brazil. The model also uses census population from the year 2000 \cite{censo}, following age group proportions. The agents are then bounded into families and they age following official mortality probability data. Female agents give birth according to their fertility rates by age and state. Number of firms and urban rural proportions also follow actual numbers and spatial coordinates. A detailed description of the model is available in \cite{seal}.

\subsection{Agents}\label{agents}

\subsubsection{Citizens, families and households.}

Citizens are bounded into families and are distinguished by gender, age and qualification. Families consolidate every working members' income and make spending decisions together. They also move and participate in the housing market as a single entity. 

Houses are fixed and can be either empty or occupied although they always have a single owner. Household addresses are the reference for all members of the family. They are relevant when searching for a job, to compute commuting statistics for those employed and when choosing firms to buy goods. Citizens, families and households each have their own class instances. Children are born and remain in their mothers' families. The model does not contemplate neither new families nor marriage at this time.

\subsubsection{Firms and government.}

Firms participate in the labor market and maintain a number of workers that varies in size according to sales performance. Firms are fixed in space and pay taxes to the government they belong to, given their location. Production is proportional to labor force and its qualification. Workers' productivity is adjusted by an $\alpha$ parameter. Technology is fixed, products are homogeneous and prices depend on the level of firms' stock \cite{blinder,dawid_2014}. 

Governments - as municipalities or as a single metropolitan region - simply collect consumption taxes at the moment of sales transaction at the location of the firm. Municipalities' QLI starts at the same level of Human Index Developments in 2000. Then, they are incrementally raised as a linear value of collected taxes within a given municipality weighted by current population. That implies a municipality with a proportionally more active goods market and smaller population collects more taxes and can increase QLI faster. Higher QLI makes houses on that municipality more expensive. 

\subsection{Markets}

\subsubsection{Goods market.}

Families consume a given probabilistic share of their current available cash depending on a propensity to consume parameter $\beta$. Families then search either for closest firm or the cheapest price in a parameter-defined sample of the market. 

Once selected the firm, the family will spend all designated money for consumption, given that the firm has enough product to supply. \footnote{Probabilistically, on average, we assume that consumption at a single shop would yield similar results if the families chose more places to shop.} Otherwise, the family consumes whatever quantity is available. The money not set to consume plus any amount not spent is saved by the family. Such saved money is afterwards reserved for use in the housing market. Taxes are paid to the local government of the firm when the family makes the purchase. 

\subsubsection{Housing market.}

House prices are given by a fixed size, intrinsic quality and the QLI of their region \cite{malpezzi}. The QLI represents amenities of the municipality and richer municipalities have higher QLI levels. The housing market follows empirical results for the Brazilian case which suggest that there are always unoccupied houses \cite{nadalin,retrato}. Every month a percentage of families enter the housing market. All vacant houses are also available for purchasing. 

Both families and houses are sorted. Cheaper houses and families with less savings rank first. If the family has enough savings to purchase a house, the transaction happens. If not, the savings of the next family are checked. When there are no more families with enough funds or available houses, the market closes. 

Every time a family buys a new house, they will evaluate their option of actually moving to the new house. If the family lives in their best house, but all adults are unemployed, they move to the second best house and then make the best house available for sale in the market the following month. On the contrary, if the family does not live in the best house and at least one adult is employed, they move into the best house the family owns. In all other cases, the family does not move. 

Such a system of moving into the second best house and selling the best-quality one, slowly tends to make the poorest move to the outer municipalities where housing is cheaper \cite{brueckner}. Moreover, even though it would be in the family's interest to remain in the affluent area, they actually are unable to afford it and need to get liquidity to participate in the consumption market. 

\subsubsection{Labor market.}

All agents that are not employed and are over 15 or below 70 years old enter the labor market every month. Firms evaluate a number of strategies. They quantify a 'cash reserve' value which enables them to pay out employees before laying them off. If the firm's available cash is above such cash reserve threshold, the firm enters the labor market hiring.

Further, if the firm is currently with no employees but either their stock of products or available cash is positive, they also enter the market. When the firms' current cash drops below the monthly payroll, they start firing employees, one at a time.

Once offering firms and applicants are defined, the labor market operates under two alternatives: half the applicants are chosen according to their qualification and half given their proximity to the firms' address. This is relevant for the case of Brazil as commuting is expensive and time consuming due to its low quality general inefficiency. Further, employers are legally bind to pay employees' commuting that exceeds 6\% of their salary in what is called 'commuting-voucher'. In practice, that means business discriminates for place of residence for non-specialized positions.  

Firms that had profit in the last quarter and have available cash above their 'cash reserve' level distribute such surpluses to their employees proportionally to their qualification level. 

Finally, firms are sorted according to their current calculated paying wages. Firms paying higher salaries choose first and get employees with better qualification or who live closer to their firms. 

\subsubsection{Spatial and markets dynamics.}

Those rules of the markets provide a very dynamically spatial model with families moving within and across municipalities. Firms getting varying demand and labor offers and municipalities collecting ever-changing taxes which affect their QLI at different speeds. When QLI increases too much, houses get more expensive and less families can afford to move in. Less populated municipalities then may sell less which in turn decreases tax collection. 

\subsubsection{Parameters}

This model has not been fully parametrized to fit a real example. However, as made explicit in \ref{limitations}, a general economic pathway in terms of macro indicators is observable. A brief sensitivity analysis of the model was conducted in order to check whether small changes in the parameters alters the general results obtained. The simulation of this paper uses the following parameters: $\alpha$ = .3, $\beta$ = .94, percentage of families entering the real estate market = .004, vacant houses percentage = .09, and taxes are set at .25. Distance as a hiring criteria is also a parameter.

\subsection{Scheduling and multiple simulation runs}

\subsubsection{Simulation runs} 

The model is structured so that a single simulation or multiple runs can be made automatically. That means either the model runs once from 2000 to 2020 and the outputs are produced or the model repeatedly runs for a given number of times and the accumulated results are produced. In each run, different random numbers are used. That influences many aspects of the model, such as: the choice of amount for families' consumption, entering the real estate market, demographic probabilities, chance of being fired, among others.

\subsubsection{Scheduling}

Agents are generated and saved for a given sample of population. In a single run, the model first saves or loads the agents (\ref{agents}). Then actions are implemented in a day, month, quarter and year pattern. The only activities that happen in business days (set to 21) are daily production and commuting. Monthly activities concentrate the core of the model and happen in the following order:

\begin{enumerate}
	\item Process demographics (aging, births, deaths).
	\item Firms make payments.
	\item Families redistribute their own cash within members.
	\item Family members consume (goods market).
	\item Governments collect taxes.
	\item Governments spend the collected taxes on life quality improvement.
	\item Firms calculate profits and update prices.
	\item Labor market is processed.
	\item Real estate market is processed.
	\item Statistics and output are processed.
\end{enumerate}

\section{Results}\label{results}

\begin{figure}[t!]
	\centering
	\includegraphics[width=12cm]{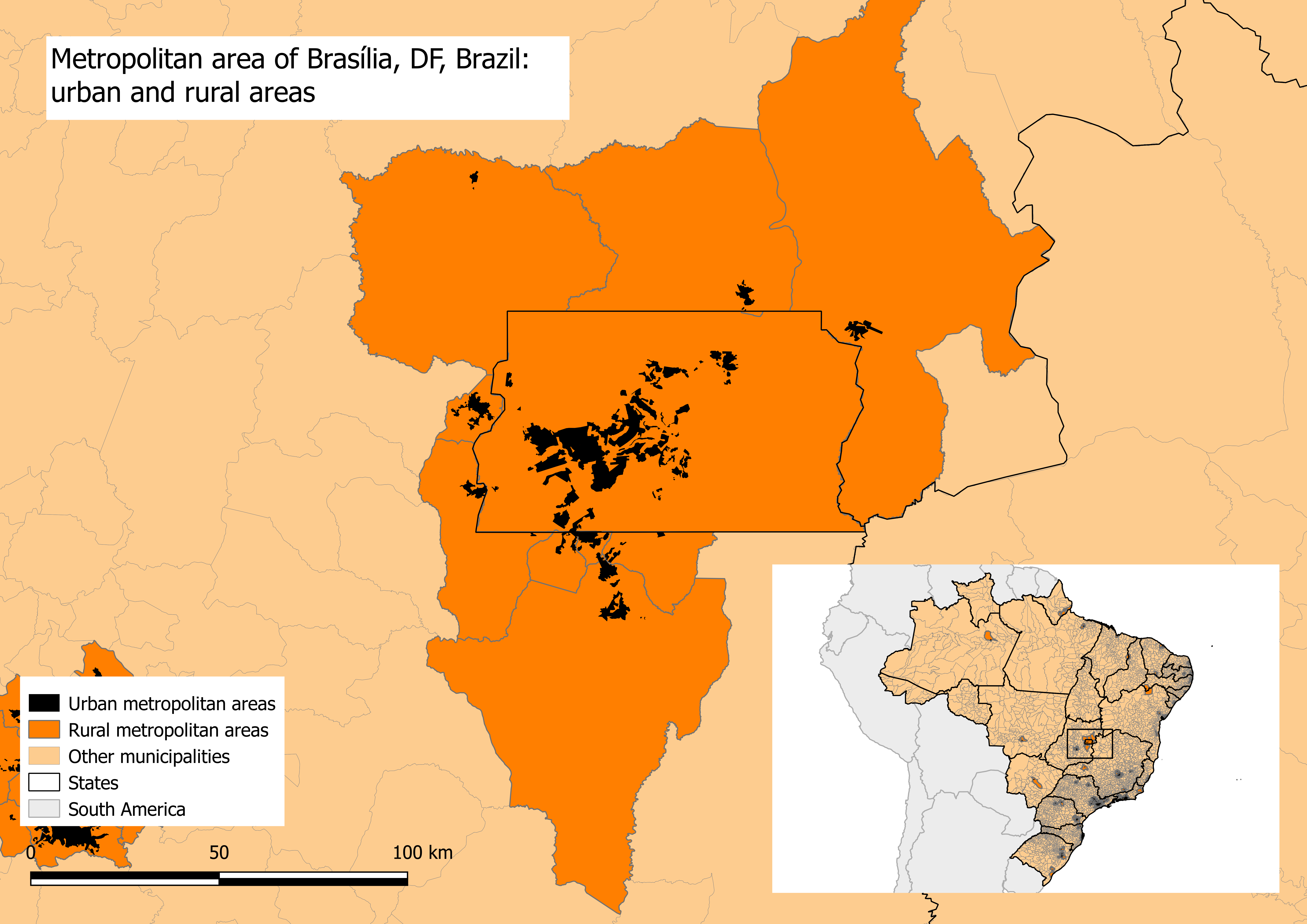}
	\caption{Case study: Bras\'{i}lia. The 10 municipalities in orange officially constitute the metropolitan region (RIDE). The run of the  individual municipalities are compared with the run of the model considering the full RIDE as one single government entity. Urban areas, in black, are proportionately more populated. Commuting (Euclidean distance) is considered among households and firms in both labor and goods market.}
	\label{fig1}
\end{figure}

Bras\'{i}lia and its metropolitan region (RIDE) with 10 municipalities is the illustrative case study (see Figure \ref{fig1}). In order to evaluate the results, a simulation from 2000 to 2020 was repeated a total of 20 times (with different random seeds). The same model was run 10 times with the municipalities being considered independent entities just as they are today. Another 10 runs were made as if all municipalities were one single government entity. That way, for half of the simulation runs taxes are collected locally and lead to QLI increase within the municipality where the transactions were made. For the other half, taxes are globally collected and homogeneously improve the QLI for all municipalities.

\begin{figure}[t!]
	\centering
	\includegraphics[width=12cm]{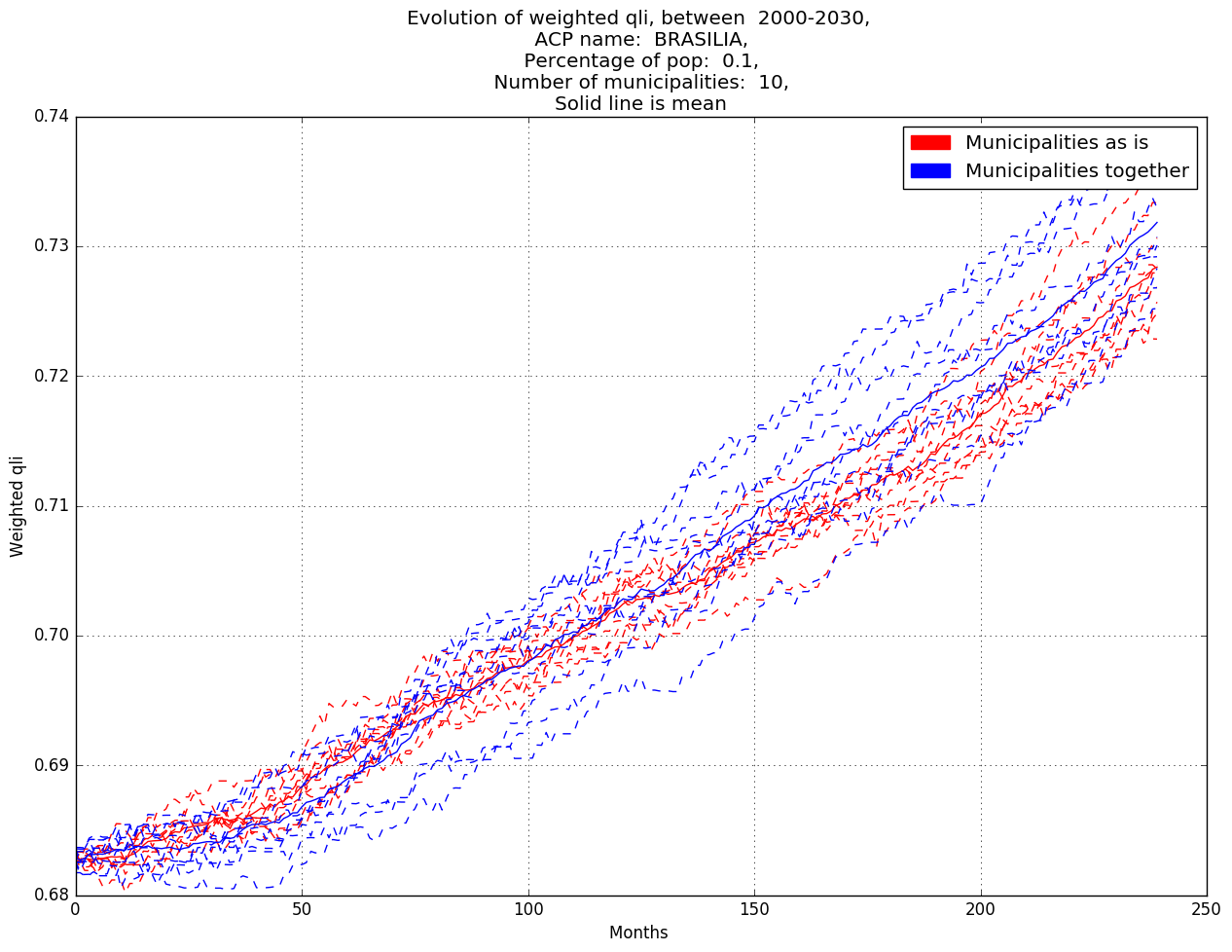}
	\caption{Results of mean weighted Quality of Life Index (QLI) for Bras\'{i}lia, 20 runs. Blue lines represent runs with one government entity. Red lines boundaries are as they are currently: 10 individual municipalities. Solid lines are mean values. QLI increases according to government revenue collected for a given municipality. Housing prices are influenced by QLI at their locations.}
	\label{fig2}
\end{figure}

Results indicate that, given the implicit variability of the model, average population weighted QLI follow similar trajectory (see Figure\ref{fig2}). Actually, a $t$ test could not reject the hypothesis of identical averages between the two alternatives (together and separated municipalities)at 5\% (t-statistic=-1.742). Although the QLI is slightly superior for the runs in which the municipalities are together, we are unable to state that such result holds for other samples of the population. 

\begin{figure}
	
	\centering
	\subfloat[individual municipalities]{{\includegraphics[width=12cm]{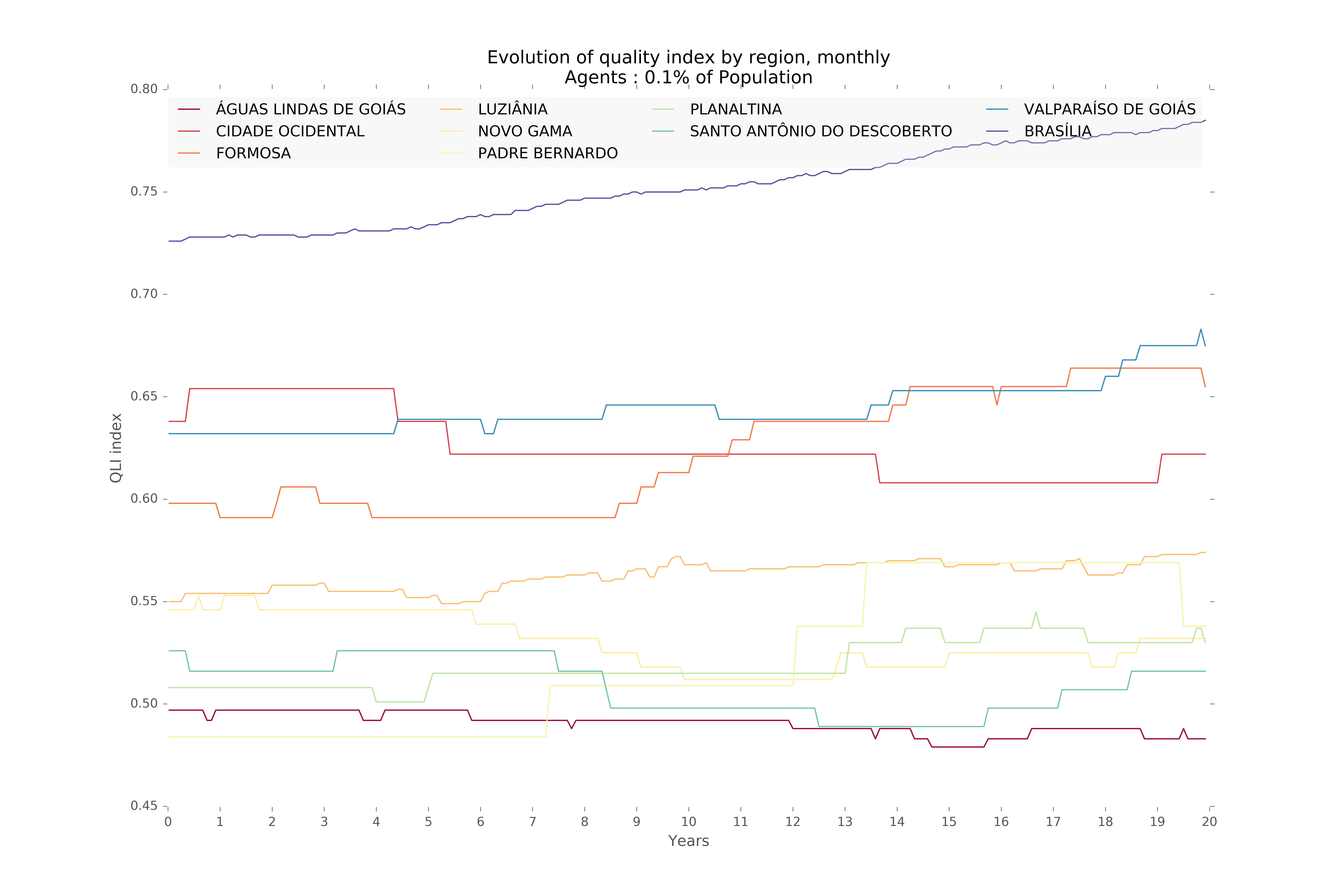}}}
	\qquad
	\subfloat[municipalities as single entity]{{\includegraphics[width=12cm]{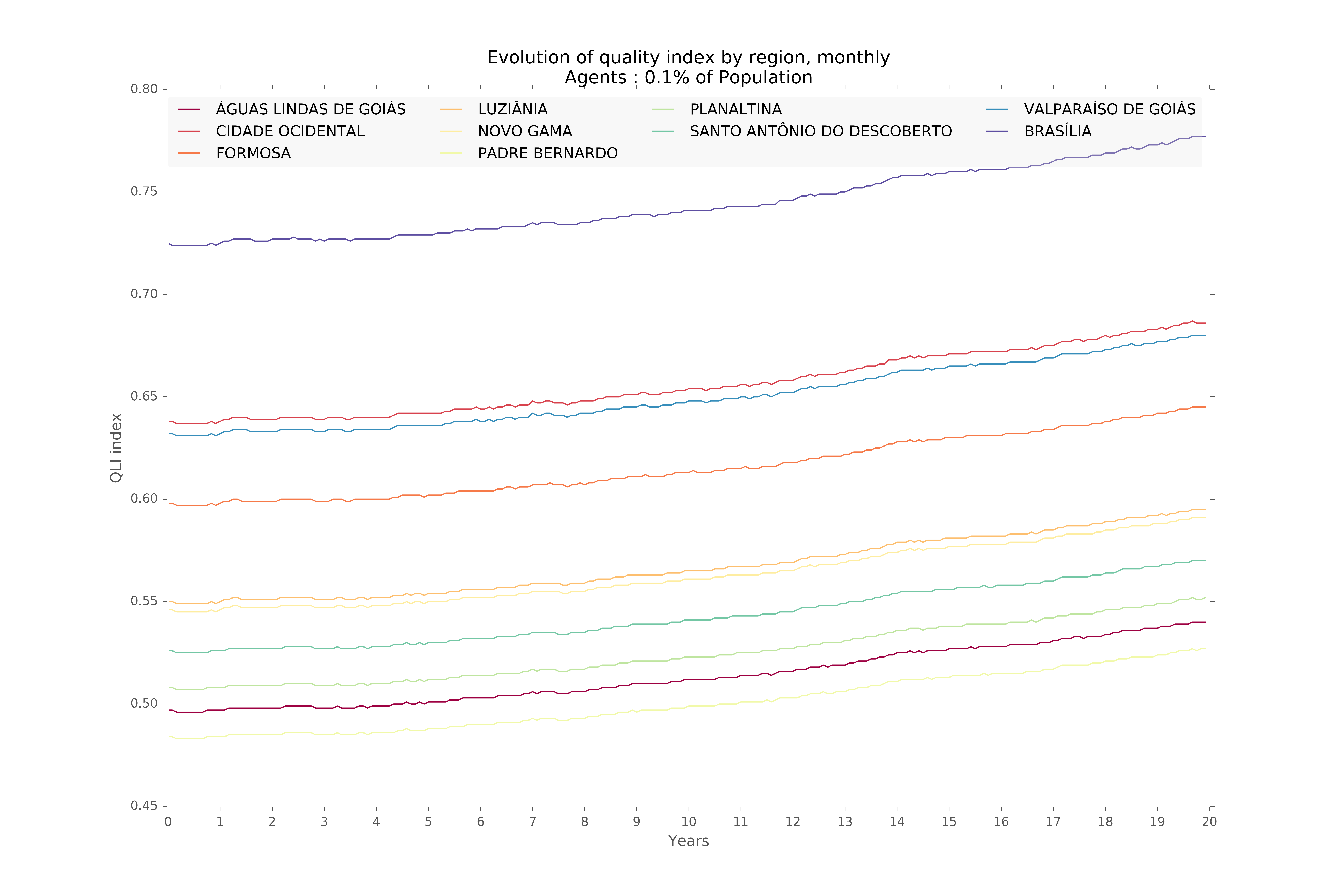}}}
	\caption{QLI evolution of each municipality. On the top, municipalities collect taxes as individual entities and invest in their own territory. On the right-hand side, all taxes collected in the region are distributed homogeneously as if they were a single government entity.}%
	\label{fig3}
\end{figure}

We are evaluating whether the increase in QLI has an absolute effect on all citizens. However, if we focus on the individual poorer municipalities with lower values for QLI, we can tell that a distributive effect actually happens with gains observed in Figure \ref{fig3}b when compared to flat lines in Figure \ref{fig3}a. Of course, such gains derive from a slower increase from Bras\'{i}lia, resulting in a very small absolute result for this model. 

Finally, these outputs of the model despite being illustrative show that it is possible to use spatially separated markets to observe differentiated behavior across the territory. In fact, although average weighted QLI is the same, its distribution in space is considerably different.

\subsubsection{Sensitivity analysis}

The sensitivity analysis indicates that the changes affected by the parameters are small in magnitude. Among the parameters tested, $\alpha$ (labor productivity), $\beta$ (propensity to consume) and Tax on consumption seem to impact the most. Alpha produced the largest GINI index at the maximum value at 0.5, but the lowest GINI was found at 0.3 and not the minimum value. Consumption of the families is also higher at the lowest level of taxes. In general, lower taxes (5\%), implied in higher levels of firms' and families' wealth as well as a higher absolute GDP value. Finally, higher values of beta lead to higher production, but also higher inequality.

Size of the market, house vacancy, percentage of families entering the real estate market or use of distance in the hiring process do not seem to impact the general trends of the results in a relevant manner. Some variation is observed in absolute GDP output as well as in the time onset of some trends. When distance is totally disconsidered and hiring criteria is restricted to qualification only, the economy seems to reach stability earlier. In contrast when distance is the sole determinant of hiring, the economy takes longer to become stable and GINI values are lower. 

\subsubsection{Limitations}\label{limitations}

The results presented here are only illustrative. In order to make them more robust, it would be necessary to run a higher number of simulations with varying samples of the population so that a trend could be identified, similarly to the proposed by \cite{anselin}. Further, hitherto we have been unable to implement multiprocessing and parallel programming while maintaining an object oriented programing paradigm and intense agents  interaction. Consequently, given the need of repeated runs with large number of agents, simulations are taking unreasonably long.

Considering the content and the investigation of administrative boundaries, we believe we would have to expand the tax systems in order to make it more similar to observed taxes. That is especially relevant because of the regressive tax system in Brazil that tends to favor metropolitan capitals \cite{furtado_finances}. 

Once a full tax system is implemented, the current model needs to be validated against observed data. As it is, unemployment remains between 0 and 5\% of the working force from the 5th year onwards. GINI coefficient levels are below Brazilian standards at 0.28 on average. GDP per capita growth varies around 0.7\%. Finally, this model, as the abstract one, does not contain a credit market. 

\section{The framework and final considerations }\label{final}

Turning a simple abstract model into an empirical one has presented some challenges. Firstly, implementing demographics dynamics meant reading age group, gender and location of population size, aging all the agents and allowing for fecundity while controlling for mortality by region.

Secondly, another difficult task is the need to maintain a system in which firms keep a permanent working force, but have to face very dynamic and erratic sales. Additionally, keeping the system on a growth path means avoiding generalized shut down of firms, constant consumption of families and an active labor market. 

Thirdly, given the nature of agent-based models and its intrinsic stochasticity, every test should include repeated simulations so that the conclusions are based on most likely scenarios. However, running simulations with a 10\% of the sample for the case of Bras\'{i}lia would mean 273,000 agents and such simulation would need to be run at least 100 in order to guarantee robust results. Finally, a sensitivity analysis with six main parameters and ten interval tests varying one parameter per time only as the others remain fixed would imply 60 further simulation runs.

Given such a context, this paper tests an illustrative \textit{ex-ante} policy analysis using simple markets and spatially bounded, local government entities. Thus, it demonstrates the viability of a stable economic framework in which further experimentation can be tried upon. In fact, SEAL's framework enables a large number of policy analysis, including, at least: (a) transportation and commuting studies, given that firms and households are known and dynamic in the case of the latter, (b) price and wage strategies at the level of the firm specifically, (c) labor market rules and relevance of workers' qualification, and (d) efficiency of municipalities' public investments.


\begin{thebibliography}{5}
	
\bibitem{ahrend}
Ahrend, R., Farchy, E., Kaplanis, I., \& Lembcke, A. C. (2014). What makes cities more productive? Evidence on the role of urban governance from five OECD countries. OECD Regional Development Working Papers, 2014(5), 33.

\bibitem{anselin}
Anselin, L. (1995). Local indicators of spatial association—LISA. Geographical Analysis, 27(2), 93-115.

\bibitem{blinder}
Blinder, A. S. (1994). On sticky prices: academic theories meet the real world. In Monetary policy (pp. 117-154). The University of Chicago Press: N. Gregory Mankiw.

\bibitem{boero}
Boero, R., Morini, M., Sonnesa, M., \& Terna, P. (2015). Agent-based models of the economy: from theories to applications. New York, NY: Palgrave Macmillan.
	
\bibitem{censo}
Brazil: Censo demogr\'{a}fico 2000: agregado por setores censit\'{a}rios dos resultados do universo. Instituto Brasileiro de Geografia e Estat\'{i}stica (IBGE) (2003).

\bibitem{brueckner}
Brueckner, J. (1987). The structure of urban equilibria: a unified treatment of the Muth-Mills model. In Handbook of Regional and Urban Economics (pp. 821-845). Elsevier Science Publishers B.V.


\bibitem{retrato}
Cavalcanti, C. B., Oliveira, C. A. P. de, Teixeira, L. M., Giustina, Y. R. D., Furtado, B. A., Krause, C. H., … Nadalin, V. G. (2016). Retrato das \'{a}reas centrais do Brasil. 

\bibitem{colander_kuper}
Colander, D., Kupers, R.: Complexity and the Art of Public Policy: Solving Society’s Problems from the Bottom Up. Princeton University Press (2014).

\bibitem{dawid_2014}
Dawid, H., Gemkow, S., Harting, P., Van der Hoog, S., Neugart, M.: Agent-based macroeconomic modeling and policy analysis: the Eurace@ Unibi model. Bielefeld Working Papers in Economics and Management. (2014).

\bibitem{dawid2}
Dawid, H. (2015). Modeling the economy as a complex system. In Modeling complex systems for public policies (pp. 169-190). Bras\'{i}lia, DF: Bernardo Alves Furtado, Patr\'{i}cia Sakowski, Marina H T\'{o}volli.

\bibitem{derhoog}
Deissenberg, C., Van Der Hoog, S., \& Dawid, H. (2008). EURACE: A massively parallel agent-based model of the European economy. Applied Mathematics and Computation, 204(2), 541-552.

\bibitem{dosi_2012}
Dosi, G., Fagiolo, G., Napoletano, M., Roventini, A.: Income Distribution, Credit and Fiscal Policies in an Agent-Based Keynesian Model. SSRN eLibrary. (2012).

\bibitem{filatova_2009}
Filatova, T., Parker, D., Van der Veen, A.: Agent-Based Urban Land Markets: Agent’s Pricing Behavior, Land Prices and Urban Land Use Change. Journal of Artificial Societies and Social Simulation. 12, (2009).

\bibitem {furtado_jasss}
Furtado, B.A., Eberhardt, I.D.R.: A Simple Agent-Based Spatial Model of the Economy:   Tools for Policy. JASSS. 19, 12 (2016).

\bibitem{seal}
Furtado, B.A., Eberhardt, I.D.R., Messa, A.: SEAL's operating manual: a Spatially-bounded Economic Agent-based Lab. arXiv:1609.03996 [cs, q-fin]. (2016).

\bibitem{furtado_finances}
Furtado, B.A., Mation, L., Monasterio, L.: Fatos estilizados das finan\c{c}as p\'{u}blicas municipais metropolitanas brasileiras entre 2000-2010. In: Territ\'{o}rio metropolitano, pol\'{i}ticas municipais. pp. 291–-312. Bernardo Alves Furtado; Cleandro Krause; Karla Fran\c{c}a, Bras\'{i}lia (2013).

\bibitem{furtado_modeling_book}
Furtado, B.A., Sakowski, P.A.M., T/'{o}volli, M.H.: Modeling complex systems for public policies. IPEA, Instituto de Pesquisa Econ\^{o}mica Aplicada, Bras\'{i}lia (2015).

\bibitem{gaffeo}
Gaffeo, E., Gatti, D.D., Desiderio, S., Gallegati, M.: Adaptive microfoundations for emergent macroeconomics. Eastern Economic Journal. 34, 441–-463 (2008).

\bibitem{geyer_cairney}
Geyer, R., Cairney, P.: Handbook on complexity and public policy. Edward Elgar Publishing (2015).

\bibitem{guocheng}
Guocheng, W., Yuna, S., Jie, W., \& Zili, W. (2015). Application analysis on large-scale computation for social and economic systems. Presented at the International Conference on Systems, Man and Cybernetics, Hong Kong: IEEE SMC.

\bibitem{lengnick}
Lengnick, M.: Agent-based macroeconomics: A baseline model. Journal of Economic Behavior and Organization. 86, 102-–120 (2013).,

\bibitem{malpezzi}
Malpezzi, S. (2002). Hedonic pricing models: a selective and applied review. In Housing Economics. Kenneth Gibb e Anthony O'Sullivan.

\bibitem{miller_page}
Miller, J.H., Page, S.E.: Complex adaptive systems. Princeton University Press (2007).

\bibitem{nadalin}
Nadalin, V., Igliori, D.: Empty spaces in the crowd. Residential vacancy in S\~{a}o Paulo's city centre. Urban Studies. 0042098016666498 (2016).

\bibitem{parker}
Parker, D. C., Manson, S. M., Janssen, M. A., Hoffmann, M. J., \& Deadman, P. (2003). Multi-Agent Systems for the Simulation of Land-Use and Land-Cover Change: A Review. Annals of the Association of American Geographers, 93(2), 314-337. https://doi.org/doi:10.1111/1467-8306.9302004

\end{thebibliography}
\end{document}